\documentclass[usegraphicx]{mn2e}

\usepackage{times}

\title[The 3-D structure of the Magellanic Clouds]{Red variables in the OGLE-II
data base -- III. Constraints on the three-dimensional structures of 
the LMC and SMC}
\author[P. Lah et al.]{P. Lah$^{1,2}$, L. L. Kiss$^2$\thanks{E-mail:
l.kiss@physics.usyd.edu.au}\thanks{On leave from University of Szeged,
Hungary} and  T. R. Bedding$^2$\\
\\
$^1$Research School of Astronomy \& Astrophysics,
Australian National University, Canberra, Australia\\
$^2$School of Physics, University of Sydney 2006, Australia}

\begin{document}

\date{Accepted ... Received ..; in original form ..}

%\pagerange{\pageref{firstpage}--\pageref{lastpage}} \pubyear{2002}

\maketitle

\begin{abstract}

We present an analysis of the 3-D structure of the Magellanic Clouds, using
period--luminosity (P--L) relations of pulsating red giants in the OGLE-II
sample. By interpreting deviations from the mean P--L
relations as distance modulus variations, we examine the three-dimensional
distributions of the sample. The results for the Large Magellanic Cloud, based
solely on stars below the tip of the Red Giant Branch, confirm previous
results on the inclined and possibly warped bar of the LMC.
The depth variation across the OGLE-II field is about 2.4 kpc, interpreted
as the distance range of a thin but inclined structure. The inclination angle is about
29$^\circ$. A comparison with OGLE-II red clump distances revealed intriguing differences 
that seem to be connected to the red clump reddening correction. A spatially variable
red clump population in the LMC can explain the deviations, which may have a broader 
impact on our understanding of the LMC formation history. For the
Small Magellanic Cloud, we find a complex structure showing patchy distribution
scattered within 3.2 kpc of the mean. However, the larger range of the overall
depth on every line-of-sight is likely to smooth out significantly the real variations.

\end{abstract}


\section{Introduction}

The close proximity of the Magellanic Clouds offers an important opportunity to
study in detail the dynamics and composition of these galaxies. One of the
important issues that affects our interpretation of the observable parameters is
the three-dimensional structure of the Clouds. The Large Magellanic Cloud  (LMC)
has long  been regarded as a thin flat disk seen nearly face-on, but  with the
east side being closer than the west (Westerlund 1997; see also  van der Marel
2004 for an updated review). Recent determinations of the tilt angle place it in
the range 30--35$^\circ$ (van der Marel \& Cioni 2001, Olsen \& Salyk 2002,
Nikolaev et al. 2004). However, one can find a significantly broader range in
the literature  (25--45$^\circ$, Westerlund 1997), which can at least partly be
explained  by nonplanar geometry of the inner LMC (Nikolaev et al. 2004). In
contrast, the structure of the Small Magellanic Cloud (SMC) has proved difficult
to define, mostly because of its large inclination angle. Different distance
indicators implied a line-of-sight depth from 7 kpc (O, B and A type stars;
Azzopardi 1982) to 6--12 kpc (star clusters; Crowl et al. 2001) and 20 kpc
(Cepheids; Mathewson et al. 1986, 1988). It is well established that the
asymmetric appearence of the SMC is due almost exclusively to the young
population of stars, since the older stellar population shows a very regular
distribution (e.g. Zaritsky et al. 2000, Cioni et al. 2000, Maragoudaki et al.
2001).  

Time-series observations from microlensing projects such as MACHO  (Alcock et al.
2000) and OGLE (Udalski et al. 1992), combined with data from two recent
near-infrared surveys, 2MASS (Skrutskie 1998) and DENIS (Epchtein 1998), have
provided an unprecedented new wave of  high-quality data and a
dramatic improvement in statistical analyses. The order-of-magnitude increase in
the numbers of well-measured stellar distance indicators has allowed detailed
searches for small distance modulus variations across the face of the Clouds.
Most recent examples for the LMC include the  Cepheid period--luminosity (P--L)
relations (Nikolaev et al.  2004), the Asymptotic Giant Branch (AGB) luminosity
distribution (van der Marel \& Cioni 2001) and the luminosity of core
helium-burning red clump stars  (Subramaniam 2003, 2004). The basic idea was the
same in all these  studies: the observed apparent magnitude deviations were
translated to distance modulus variations. The analysed fields ranged from the
central 4.5 square degrees of the LMC bar (Subramaniam 2003, 2004) to over 300
square degrees, covering the outest parts of the disk (van der Marel \& Cioni
2001).

In this Letter we present the first attempt to use red giant P--L relations to
constrain the three-dimensional structures of the Clouds. Since the
discovery of multiple P--L relations of long-period variables in the LMC  (Wood
et al. 1999, Wood 2000), there has been strong interest in pulsating red giants
and many independent analyses have been published (Cioni et al. 2001, 2003;
Noda et al. 2002, 2004; Lebzelter, Schultheis \& Melchior 2002; 
Kiss \& Bedding 2003, 2004 -- Paper I and Paper II; Ita et al. 2004ab; 
Groenewegen 2004; Schultheis, Glass \& Cioni 2004; Soszy\'nski et al. 2004).
Interestingly, none of these studies considered the geometric effects on
the apparent magnitudes, as has been done, for instance, by 
Wray, Eyer \& Paczy\'nski (2004), who analysed short-period red giants in 
the galactic Bulge. The main aim of the
present paper is to address this issue and draw conclusions on the line-of-sight
distance variations of the Clouds, as traced by pulsating red giants. 

\section{Data analysis}

In this work we used OGLE-II periods and 2MASS K$_{\rm s}$  magnitudes of red
giant variables in the LMC and SMC, as discussed in Papers I and II, where all
relevant details on data reduction can be found. Our key assumption here is that
the vertical scatter about any particular P--L relation contains information on
the relative distances to individual stars. This allows us to examine the
magnitude differences from mean P--L relations as a function of position on the
sky.  Although this sounds simple, we have to ask to what extent this
assumption is valid. There are three aspects to the question: {\it (i)}
star-to-star extinction variations due to interstellar dust;  {\it (ii)} the
contribution of stellar variability to the vertical scatter in the P--L
relations; {\it (iii)} the intrinsic width of the P--L relations.

\begin{table}  % location here, top, bottom, separate page
\begin{centering}
\caption{The inverse regression coefficients of the 
fitted P--L relations.  
They have the form $\log{P}~=~a^\prime~\times~K_S~+~b^\prime$.
`Number' refers
to the number of stars used in the line fit for that P--L relation.} 
\label{linefits}  % name for referencing 
\begin{tabular}[b]{|l|c|c|c|}  
%  c = centered input | draws vertical line
\hline % horizontal line - top of table
P--L relation$^a$ & $a^\prime$ & $b^\prime$ & Number \\ 
\hline 

  % LMC
\multicolumn{4}{|c|}{{\bf LMC }} \\ % spans 4 columns centered

R$_3$ (A$^-$) & $-0.237 \pm 0.002$  & $4.340  \pm 0.029$ & $2642$\\   

R$_2$ (B$^-$)& $-0.269 \pm 0.004$ & $4.931 \pm 0.045$ & $1634$\\   

\multicolumn{4}{|c|}{ } \\ % spans 4 columns centered

  % SMC
\multicolumn{4}{|c|}{{\bf SMC }} \\ % spans 4 columns centered

R$_3$ (A$^-$) & $-0.206 \pm 0.006$ & $4.030 \pm 0.120$ & $229$ \\

R$_2$ (B$^-$) & $-0.214 \pm 0.006$ & $4.330 \pm 0.130 $ & $117$ \\

3O (A$^+$) & $-0.206 \pm 0.006 $ & $4.038 \pm 0.070$ & $133$ \\

2O (B$^+$)& $-0.214 \pm 0.006$ & $4.349 \pm 0.069$ & $218$ \\

1O (C$^\prime$)& $-0.238 \pm 0.006$ & $ 4.821 \pm 0.069$ & $260$ \\

F (C) & $-0.222 \pm 0.007$ & $4.932  \pm 0.078$ & $405$ \\

L$_2$ (D)$^b$ & $-0.170 \pm 0.005$ & $4.799  \pm  0.062$ & $534$ \\

L$_2$ (D)$^c$ & $-0.170 \pm 0.005$ & $4.810 \pm 0.160$ & $405$ \\

\hline % bottom of table
\end{tabular}
\end{centering}
\vskip1mm
$^a$ -- abbreviations by Ita et al. (2004a) in parentheses\\
$^b$ -- above the tip of the Red Giant Branch\\
$^c$ -- below the tip of the Red Giant Branch\\
\end{table}

 {\it (i)} The influence of dust absorption in the near-infrared was discussed by
van der Marel \& Cioni (2001) for the LMC, who concluded that it was only a few
hundredths of a magnitude in $K_S$ and could be neglected. This has later been
criticised by Nikolaev et al. (2004), who preferred the multi-wavelength approach
that involved both distance modulus and extinction variations. Their fig.\ 7
shows that most of the  individual Cepheid  reddenings scattered within $\delta
E(B-V)\sim\pm0\fm1$, which can be translated to a random error $\delta
K_s\approx\pm0\fm03$, using the extinction law in Schlegel et al. (1998).
However, there are also two facts which have to be considered: 1. the Nikolaev et
al. sample covered a much larger  field of view, where larger extinction
variations can occur; 2. Cepheids, as young supergiant stars, are often found
close to star forming regions, where the local dust amount might be much higher
than the average. Individual reddenings have also been determined by Subramaniam
(2003), who analysed exactly the same OGLE-II field of view as us (Udalski, Kubiak 
\& Szymanski 1997), allowing a direct comparison. She found
$\Delta_{max}E(V-I)\approx 0\fm029$ as the maximum reddening dispersion along the
bar of the LMC, which corresponds to $\delta K_s\approx\pm0\fm01$ extinction
dispersion in $K_S$. From these numbers, we conclude that for each star, the
uncertainty in $K_S$  due to extinction is likely to be less than 0\fm03,
possibly around 0\fm01, which is indeed negligible if we can take averages over
hundreds of stars.

{\it (ii)} Since 2MASS magnitudes are single-epoch measurements, intrinsic
variability can introduce some uncertainty. For that reason, we decided to use
only small-amplitude first ascent Red Giant Branch (RGB) stars in the LMC, of
which several thousands were identified in Paper I. Their amplitudes in $I$ range
from 0\fm005 to 0\fm02, so that 2MASS $K_S$ measurements should be within
0\fm01--0\fm02 of the mean values. The situation is less favourable in the SMC,
for which the smaller number of stars (about 3,200 were analysed in Paper II)
forced us to use most P--L relations, including both RGB and AGB variables. The
latter may have amplitudes up to several tenths of a mag in the infrared
(Whitelock, Marang \& Feast 2000).

{\it (iii)} The horizontal width of the P--L relations is affected by errors 
arising from period determination and instrinsic scatter caused by astrophysics
(e.g., mixing different populations of stars or random and time-dependent
excitation of pulsations). Considering P--L ridges R$_2$ and R$_3$, we compared the relations
presented in Paper I with those determined by Wood (2000) and Soszy\'nsky et al. (2004),
both based on about 8 years of observations (twice as long as OGLE-II).
If the widths were dominated by errors in period determination caused by the short time-span 
of data, the longer datasets 
should have led to significantly tighter P--L relations. Somewhat surprisingly, 
this is not the case: we found the same $\sim$0.1 dex width in $\log P$
in all three studies, which implies that the natural width of the relations is probably 
resolved by
OGLE-II alone. We note that errors in period determination can also arise from the fact 
that the majority of stars show multiply periodic behaviour, so that period uncertainty 
does not necessarily scale inverse-proportionally with the full time-span.
In this work we assumed that the scatter in the P--L relations is 
partly due to a random error in period, which will 
be averaged out in a large sample, and that there is a vertical scatter due to
a statistical error in the distance moduli estimates.

To calculate distance modulus variations, each star in our sample was  classified as
lying on one of the P--L relations. To minimize the effects of spurious periods,
we only used the dominant period for each star. 
For the classification, we drew
straight line boundaries between the different  relations, similarly to Ita et
al. (2004b). The results do not depend heavily on the exact definition of these
lines because the majority of the stars lie away from the boundaries. To
distinguish between AGB and RGB stars, we adopted TRGB magnitudes from Paper II.
We then made linear least-squares fits to each of the P--L relations. Because of
the non-zero width and the rhomboid-shaped P--L ridges, these fits were performed
as inverse regressions in the form of $\log~P=a^\prime\times K_S+b^\prime$ 
(Table\ \ref{linefits}), which 
were then converted back to usual P--L relations in
form of $K_S=a\times\log~P+b$. For a star at a given period, the
vertical difference between its observed $K_S$ magnitude and the linear fit was
taken as the distance modulus of that star relative to the average distance of
the host galaxy. Because of the relatively  large scatter in these distance
moduli, caused by the issues discussed above, it was helpful to bin the distance
data to see the underlying trends. When combining the results from different P--L
relations, each measured distance modulus was weighted by the overall scatter
(combined with the number of points) in its P--L relation.  This gave more weight
to the distances based on the P--L relations that had the tightest overall
correlations.

\section{Results}

\subsection{Structure of the LMC}

The results presented here for the LMC used only the R$_3$ and R$_2$
period-luminosity relations from Paper I, i.e. those below the TRGB (sequences A$^-$ and
B$^-$ in Ita et al. 2004a). They had by far the tightest relations and gave the
best distance measurements; 4,276 stars were used for the LMC.  

\begin{figure}
\begin{center}  % centered
\leavevmode  % fixes centering problem in figure
  \includegraphics[width=85mm]{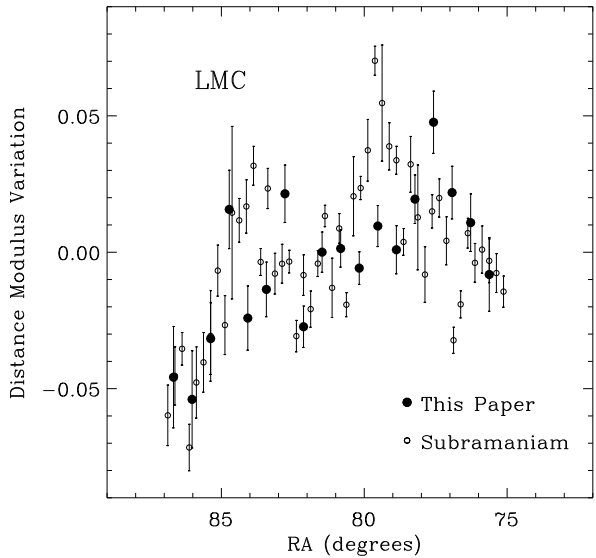}
\end{center}
\caption{The binned $K_S$ magnitude differences for various Right Ascensions, compared with
red clump results by Subramaniam (2003). Larger values correspond to those parts of the 
LMC that are further from us.}
\label{LMC_Sub}
\end{figure}

The OGLE-II data are limited to the region around the bar of the LMC, which lies approximately
in the east-west direction. Therefore, the obvious way to bin the data was in Right Ascension.
This is the same approach as used by Subramaniam (2003, 2004), who analysed
dereddened OGLE-II red-clump magnitude  variations in terms of relative distance variations.
She has kindly provided her data and we compare these with our results in Fig.\ \ref{LMC_Sub}.
To overlay  the Subramaniam $I_0$ distance moduli, we subtracted their weighted mean value
(($\langle I_0 \rangle=18\fm16$). Larger values in Fig.\ \ref{LMC_Sub} refer to those parts of
the LMC that are  further from us. The uncertainty in each RA bin was calculated from the
scatter of the distance moduli that were combined to make that point.  

It is clear from Fig.\ \ref{LMC_Sub} that red giant P--L relations alone
reveal that the bar is inclined away from us,
running east-west. This is in perfect agreement with the general view of the 
LMC (Westerlund 1997). The magnitude variations can be translated to
distance variations by assuming a distance modulus to the LMC of 
$(m-M)_0=18\fm5$ (Alves 2004a), corresponding to 50.1 kpc mean distance. The
maximum distance modulus range we observe in the LMC is 0\fm1$\pm$0\fm03
(calculated from a linear fit between RA=77$^\circ$ and 87$^\circ$),
which can be translated to 2.4$\pm$0.7 kpc distance variation. We interpret this 
as the distance range of a thin but inclined structure, where 
the thickness of the LMC bar is assumed to vary less than the mean distance along any
given line-of-sight.
Using Fig.\ \ref{LMC_Sub} and the assumed LMC distance, an inclination 
angle of 29 degrees (with a formal error of several degrees) can be determined.
However, note that OGLE-II
data are not well suited for inclination angle determination because the field
of view is highly linear so the position angle of the node,
along which the inclination angle has to be measured, is weakly constrained. In
this sense, the inclination angle we determined is a lower limit to the real
value. Nevertheless, our 29$^\circ$ is in good agreement with recent inclination
angle determinations (e.g., van der Marel \& Cioni 2001, Nikolaev et al. 2004). 

\begin{figure}
\begin{center}  % centered
\leavevmode  % fixes centering problem in figure
  \includegraphics[width=8.5cm]{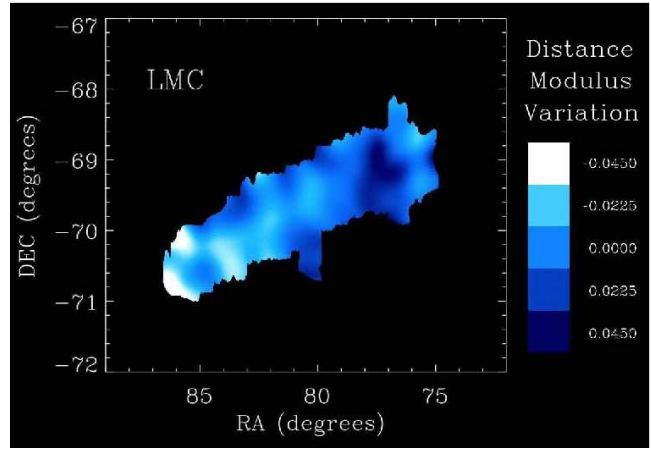}
\end{center}
\caption{A 3-D representation of the LMC. The lighter 
regions are closer to us and the darker regions are further away.}
\label{LMC_3D}
\end{figure}

There is also evidence for substructures within the  bar as deviations from a
straight line in Fig.\ \ref{LMC_Sub}. Similar substructures were recently
found by Subramaniam (2003, 2004), who interpreted red clump
magnitude  variations as evidence of a misaligned secondary bar within the
primary one. There appears to be good consistency between the two data
sets, with the distance values mostly agreeing to within  1$\sigma$. 
However, there are also some discrepancies, most prominently at 
RA$\sim 79^\circ$ and RA$\sim83-85^\circ$. At these position there are two $\sim$0\fm05 dips in
the Subramaniam data but not in our data (although we have fewer data points, so that the
2-3 $\sigma$ differences may arise purely from statistical fluctuation). 
There are two reasons to believe that our data in these regions are to be preferred. 
Firstly, optical and narrow-band 
H$\alpha$ images show prominent HII regions in the quoted positions 
(e.g. giant shells No. 51, 54 and 60 at RA$\sim79^\circ$ and No. 77 at RA$\sim84^\circ$ 
in Kim et al. 1999), which suggests that the difference may indicate
undercorrected extinction in the $I$-band data used by Subramaniam (2003). Secondly,
the two RAs where we found the largest deviations coincide exactly with the positions 
where the average $E(V-I)$ reddening values show sudden jumps in fig.\ 2 of 
Subramaniam (2003). Both facts imply that the red clump method (developed by Olsen \& 
Salyk 2002
and used by Subramaniam 2003) may be affected by a systematic error that is related to the
red clump reddening determination.

The red clump method assumes the red clump has constant $(V-I)_0$ colour everywhere in the LMC. Since we
used reddening-insensitive $K$-band data for red giants, the discrepancy with clump results
implies the intrinsic $(V-I)_0$ colour of the clump must be redder than the assumed constant
value ($(V-I)_0$=0\fm92, Olsen \& Samolyk 2002), and this probably shows a change in the mean
red clump population. Alves (2004a) showed that population effect on model red-clump
colour-magnitude diagrams can significantly affect distances derived from $V$ and $I$ band red
clump data. Our findings can be explained by a spatially varying clump population in the LMC,
which may have broad implications for understanding the LMC bulge/halo/disk structure and
formation history, as discussed recently by Alves (2004b) and Zaritsky (2004).

To search for structure in the Declination direction we created a 3-D
representation of the LMC in which the relative distance modulus is
shown in grey scale (Fig.\ \ref{LMC_3D}). This figure was made using  a
two-dimensional averaging with a Gaussian weight-function ($FWHM=8\farcm6$).
This representation makes the depression at RA$\sim77^\circ$ quite prominent. 
The overall appearance shows the tilt of the bar, with no measurable
systematic trend along Declination within the OGLE-II field of view. 

\subsection{Structure of the SMC}

\begin{figure}
\begin{center}  % centered
\leavevmode  % fixes centering problem in figure
  \includegraphics[width=85mm]{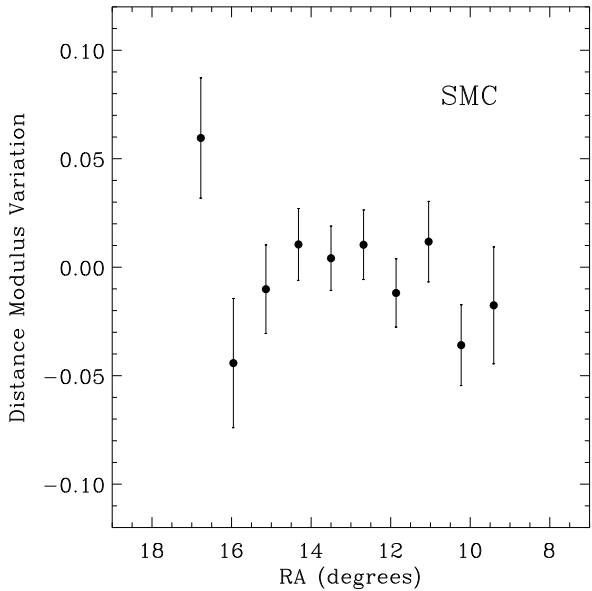}		
\end{center}
\caption{The binned $K_S$ magnitude differences for various Right Ascensions. 
The larger values refer to those parts of the SMC that are further from us.}
\label{SMC_RA}
\end{figure}

In the SMC the smaller number of stars led us to use more P--L relations to
reduce the errors. We included P--L ridges R$_3$, R$_2$, 3O, 2O, 10, F and L$_2$
(A$^-$, B$^-$, A$^+$,  B$^+$, C$^\prime$, C and D in Ita et al. 2004a); these
were the only relations for which reasonable boundaries could be defined, due to
the larger scatter in  the SMC. As for the LMC, we weighted the results by
the corresponding scatter in each P--L relation, so that better-defined
relations received higher weight. 

The resulting distance modulus variations are plotted in Fig.\ \ref{SMC_RA}. The
uncertainties are considerably larger than for the LMC, which is partly due to
having fewer stars and partly due to the known larger depth of the SMC (cf. Sect.\
1).  The latter effect is particularly prominent in fig.\ 9 of Ita et al.
(2004a): all ridges, without exception, are
significantly thicker, including those of the fundamental and first overtone
Cepheids. For that reason it would be misleading to conclude from  Fig.\
\ref{SMC_RA} that no relative distance modulus variations occur  in the SMC.
There are hints of substructures but their interpretation is difficult
because the mean distance modulus is a complicated function of
stellar distribution along the line of sight. Formally, the given distance
modulus variations correspond to a distance range of 3.2$\pm$1.6 kpc,
adopting $(m-M)_0=18\fm94$ (Paper II), which, contrary to the LMC, is likely
to be more affected by the overall depth on every line-of-sight. A better 
depth measurement would require
a study of the vertical scatter along individual P--L ridges, but for that
purpose, these red giant P--L relations in the SMC are less useful because they
are too close  to each other in the P--L plane, which makes it more difficult
to distinguish between  members of different ridges.

\begin{figure}
\begin{center}  % centered
\leavevmode  % fixes centering problem in figure
  \includegraphics[width=8.5cm]{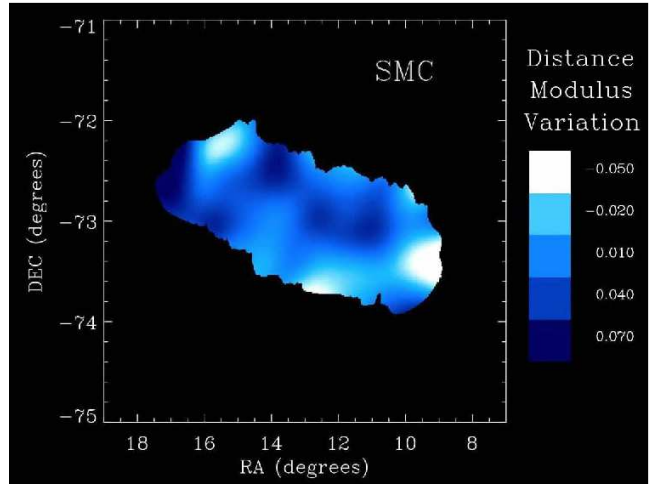}
\end{center}
\caption{A 3-D representation of the SMC. The lighter regions are closer 
to us and the darker regions are further away.}
\label{SMC_3D}
\end{figure}

Finally, the 3-D representation of the SMC (Fig.\ \ref{SMC_3D}) shows that both
the north-eastern and the south-western corners of the OGLE-II field are
slightly  closer to us, while there is a system of depressions along the
northern  boundary. Interestingly, the two deepest ``holes'' (at
RA$\sim$14$^\circ$ and 11$^\circ$) coincide exactly with the two major
concentrations of red giant (RGB and AGB) stars in figs.\ 6-7 in Cioni et al.
(2000), thus they represent significant extensions of the main body of the SMC,
inclined further away from us.

\section{Conclusions}

We have demonstrated the usefulness of pulsating red giants as distance
indicators in external galaxies by measuring the 3-D structures of the Magellanic
Clouds. In particular, the variables below the tip of the Red Giant Branch are
numerous and their narrow P--L relations are real competitors to those of the
Cepheid variables. About a year  of continuous monitoring gives very well-defined
P--L relations for those stars, which have periods between 15 and 50 days.
Because of the much lower amplitudes, intrinsic variability introduces
negligible vertical scatter in the relations. This is, of course, a major
observational obstacle: one has to reach the several millimagnitude precision
over the whole photometric monitoring period, which should be as continuous as
possible. For that reason, red giant P--L relations in general can be considered
as giving supplementary information to other distance indicators, although
high-amplitude red giants have already been used for measuring a stand-alone
extragalactic distance to NGC~5128=Cen~A (Rejkuba 2004).  

From an analysis of almost 4,300 RGB variables in the LMC we determined spatial
depth of the LMC bar as a function of celestial position. The results are in
good agreement with those based on other distance indicators (Cepheids, red
clump stars, AGB luminosity distributions), but are perhaps less affected by dust.
We also found possible evidence for spatial variations in the red clump population 
in the LMC,
which suggests that the assumption of constant $(V-I)_0$ red-clump colour over 
the whole LMC may not be valid.
For the Small Magellanic Cloud, we found a patchy structure with the two
major concentrations of late-type stars displaced a few ($<3$) kiloparsecs
further away. However, the vertical widths of red
giant P--L relations seem to be significantly higher in the SMC, which can be
attributed to the larger spatial depth range of the galaxy. Because of this, the
calculated depth change is likely to be seriously affected and the
interpretation of the results should be made with care.

\section*{Acknowledgments} 

This work has been supported by the OTKA Grant \#T042509 and the Australian Research Council.
P. Lah received a Vacation Scholarship for this project from the School of Physics, 
University of Sydney. L.L. Kiss is supported by a University of Sydney Postdoctoral Research
Fellowship. Thanks are due to an anonymous referee for very helpful comments
and suggestions. 
We also thank Dr. A. Subramaniam for providing her red clump distance data for  the LMC. 
This research has made use of the NASA/IPAC Infrared Science Archive, which is operated by the
Jet Propulsion Laboratory, California Institute of Technology, under contract with the National
Aeronautics and Space Administration.  The NASA ADS Abstract Service was used to access data
and references.

\end{document}